\documentclass[reqno,12pt,letterpaper]{amsart}
\usepackage{enumerate}
\usepackage{amsfonts}
\usepackage{amsmath}
\usepackage{amssymb}
\usepackage{amsthm}
\usepackage{latexsym}
\usepackage{xcolor}
\usepackage{mathtools}
\usepackage[pagebackref, hidelinks, bookmarks, hypertexnames=false]{hyperref}
\usepackage{amsaddr}

\calclayout

\let\varepsilonilon=\varepsilon 

\newtheorem{theo}{Theorem}
\newtheorem{defi}{Definition}

\newtheorem{corr}{Corollary}
\newtheorem{rem}{Remark}

\newcommand{\ket}[1]{\vert #1 \rangle} 
\newcommand{\bra}[1]{\langle #1 \vert} 
\newcommand{\kb}[2]{\left| #1 \vphantom{#2} \right>\left< #2 \vphantom{#1} \right|} 
\newcommand{\proj}[1]{\kb{#1}{#1}} 

\def\H{\mathcal{H}}

\def\S{\mathfrak{S}}

\def\C{\mathfrak{C}}
\def\T{\mathfrak{T}}

\def\eps{\varepsilon}
\newcommand{\supp}{\mathrm{supp}}

\newcommand{\Tr}{\mathrm{Tr}}

\newcommand{\shs}{\hspace{1pt}}
\newcounter{defin}  \newcounter{lemma}  \newcounter{theorem}
\newcounter{proposition} \newcounter{corol}  \newcounter{remark} \newcounter{example}

\newenvironment{example}{\par\refstepcounter{example}     \textbf{Example \theexample.}}{\rm\par}

\newcommand{\ha}{H} 
\DeclareMathOperator{\tr}{Tr}

\begin{document}

\title[]{Optimal continuity bound for the von Neumann entropy under energy constraints}

\author{Simon Becker}
\address{ETH Zurich, Institute for Mathematical Research, R\"amistrasse 101, 8092 Zurich, Switzerland.}

\author{Nilanjana Datta}
\address{Department of Applied Mathematics and Theoretical Physics, University of Cambridge, United Kingdom.}

\author{Michael~G. Jabbour}
\address{SAMOVAR, T\'el\'ecom SudParis, Institut Polytechnique de Paris, 91120 Palaiseau, France\\
\medskip
Centre for Quantum Information and Communication, \'Ecole polytechnique de Bruxelles, CP 165/59, Universit\'e libre de Bruxelles, 1050 Brussels, Belgium}

\author{Maksim~E. Shirokov}
\address{Steklov Mathematical Institute,\\
\medskip
Moscow Institute of Physics and Technology, Moscow, Russia\vspace{1cm}}

\begin{abstract}
Using techniques proposed in [Sason, IEEE Trans. Inf. Th. \textbf{59}, 7118 (2013)] and [Becker, Datta and Jabbour, IEEE Trans. Inf. Th. \textbf{69}, 4128 (2023)], and based on the results from the latter, we construct a globally optimal continuity bound for the von Neumann entropy.
This bound applies to any state under energy constraints imposed by arbitrary Hamiltonians that satisfy the Gibbs hypothesis.
This completely solves the problem of finding an optimal continuity bound for the von Neumann entropy in this setting, previously known only for pairs of states that are sufficiently close to each other. Our main technical result, a globally optimal semicontinuity bound for the von Neumann entropy under general energy constraints, leads to this continuity bound.
To prove it, we also derive an optimal Fano-type inequality for {random variables with a countably infinite alphabet and }a general constraint, as well as optimal semicontinuity and continuity bounds for the Shannon entropy in the same setting. In doing so, we improve the results derived in [Becker, Datta and Jabbour, IEEE Trans. Inf. Th. \textbf{69}, 4128 (2023)].
\end{abstract}

\maketitle

\thispagestyle{empty} 

\section{Introduction}\label{sec1}

Quantitative analysis of the modulus of continuity of information-theoretic quantities characterizing quantum systems is an important area of research in quantum information theory.  A comprehensive overview of key results in this domain can be found in \cite{QC,Capel}.
Among the various characteristics of quantum states, the von Neumann entropy $S$, defined in Section 2, is particularly important. Fannes \cite{Fannes} introduced the first uniform continuity bound for the von Neumann entropy on finite-dimensional spaces.
This continuity bound was subsequently tightened. Ultimately, Audenaert established an optimal (sharp) continuity bound for the von Neumann entropy of states in finite dimensions \cite{Aud}.

Winter provided the first quantitative continuity result on the von Neumann entropy for states of infinite-dimensional quantum systems \cite{W-CB}. Since in infinite dimensions, the von Neumann entropy is neither continuous nor finite across the entire set of states, the study of continuity has to be restricted to certain subsets of quantum states. Wehrl's classical work \cite{W} showed that the von Neumann entropy is continuous on the compact\footnote{The compactness of this set is proved in \cite{H-c-w-c}.} set of energy-constrained states
\begin{equation*}
\C_{H,E}=\left\{\rho\in\S(\H)\,\vert\,\Tr H\rho\leq E\right\},
\end{equation*}
where $H$ is a positive operator (Hamiltonian) satisfying the Gibbs hypothesis (condition (\ref{H-cond}) in Section~\ref{sec2}) and $\S(\H)$ is the set of states on the Hilbert space $\H.$
Consequently, finding explicit estimates on the modulus of continuity of the von Neumann entropy on this set became a natural objective.

In \cite{W-CB}, Winter established the first two faithful\footnote{That is, the right-hand side of the continuity bound tends to zero as the distance between the states in question tends to zero.} (uniform) continuity bounds for the von Neumann entropy of states within the set $\C_{H,E}$, providing accurate estimates of the modulus of continuity of $S$ on $\C_{H,E}.$ One of these bounds is asymptotically tight. Subsequent efforts aimed at improving Winter's bounds in specific cases or for specific Hamiltonians $H$ have been made (see \cite{LCB,BDJ,10849634,QC}). The primary goal of these works was to derive optimal continuity bounds for the von Neumann entropy, that is, sharp estimates of the modulus of continuity of $S$ on $\C_{H,E}.$

A partial solution was provided in \cite[Theorem~6]{BDJ}, see also \cite{10849634}, where a sharp estimate of the modulus of continuity of $S$ on $\C_{H,E}$ was obtained for any positive operator $H$ that satisfies the Gibbs hypothesis. However, this estimate is valid only for sufficiently close states $\rho$ and $\sigma$ in $\C_{H,E}$, not for all states in the set.

Initially, our goal was to find an estimate of the modulus of continuity of $S$ on $\C_{H,E}$ that are sharp for any upper bound $\varepsilon \in (0,1)$ on the trace distance of states $\rho$ and $\sigma$ in $\C_{H,E}.$ The broader aim of this research, however, was to derive a globally optimal semicontinuity bound for the von Neumann entropy, where the energy constraint is imposed on only one of the two states. Specifically, we aim to find an upper bound on the quantity $$
\Delta(\eps)=\sup\{S(\rho)-S(\sigma)\,|\,\rho,\sigma\in\S(\H),\,\Tr H\rho\leq E,\,\textstyle\frac{1}{2}\|\rho-\sigma\|_1\leq\eps\},
$$
which is sharp for any $\eps\in[0,1].$ From a technical perspective, semicontinuity bounds offer several advantages:
\begin{itemize}
\item They provide flexible continuity bounds for ``nonsymmetrical" cases (see \cite[Section 4.1]{LCB} for details);
\item They allow estimates when the energy of one of the states is unknown (see the proof of the converse claim in Theorem~1 of~\cite{Lami-new}).
\end{itemize}

\bigskip

\noindent \textbf{Outline of the remainder of the paper and summary of the results}:
\begin{itemize}
    \item In Section~\ref{sec2}, we begin with some preliminary facts about infinite-dimensional quantum systems subject to constraints imposed by Hamiltonians satisfying the so-called Gibbs hypothesis;
    \item In Section~\ref{sec:Main}, we present our two main results: (i) A semicontinuity bound for the von Neumann entropy of energy-constrained systems (Theorem~\ref{SCB}); (ii) A continuity bound for the von Neumann entropy of energy-constrained systems (Corollary~\ref{SCB-c});
  \item In Section~\ref{sec:proof}, we prove our main results. To do so, we first introduce two intermediate "classical" results in Section~\ref{sec:classRes}: (i) A Fano-type inequality for random variables with a countably infinite alphabet under a general constraint (Theorem~\ref{theo:FanoG}) and; (ii) Constrained   semicontinuity and continuity bounds for the Shannon entropy of random variables with countably infinite alphabets (Theorem~\ref{theo:classical} and Corollary~\ref{Sh-CB}). We prove 
    Theorems~\ref{theo:FanoG} and~\ref{theo:classical} 
  in Sections~\ref{sec:proofC1} and~\ref{sec:proofC2}, respectively. Those results in turn lead to the semicontinuity and continuity bounds for the von Neumann entropy, as shown in Section~\ref{sec:proofsubMain}.  \end{itemize}

\section{Preliminaries}\label{sec2}

Let $\mathcal{H}$ be a separable Hilbert space,
$\mathfrak{B}(\mathcal{H})$ the algebra of all bounded operators on $\mathcal{H}$ with the operator norm $\|\cdot\|$ and $\mathfrak{T}( \mathcal{H})$ the
Banach space of all trace-class
operators on $\mathcal{H}$  with the trace norm $\|\!\cdot\!\|_1$. Let
$\mathfrak{S}(\mathcal{H})$ be the set of quantum states (positive semidefinite operators
in $\mathfrak{T}(\mathcal{H})$ with unit trace) \cite{H-SCI,Wilde}.
The \emph{von Neumann entropy} of a quantum state
$\rho \in \mathfrak{S}(\H)$ is defined by the formula
$S(\rho)=\operatorname{Tr}\eta(\rho)$, where $\eta(x)=-x\log x$ if $x>0$
and $\eta(0)=0$. It is a concave lower semicontinuous function on the set~$\mathfrak{S}(\H)$ taking values in~$[0,+\infty]$ \cite{H-SCI,Wilde,W}.

Let $H$ be a positive (semidefinite) operator on a Hilbert space $\mathcal{H}$.  For any positive operator $\rho\in\T(\H)$ we will define the quantity $\Tr H\rho$ by the rule
\begin{equation*}
\Tr H\rho=
\left\{\begin{array}{l}
        \sup_n \Tr (P_n H\rho)\;\; \textrm{if}\;\;  \supp\rho\subseteq {\rm cl}(\mathcal{D}(H))\\
        +\infty\;\;\textrm{otherwise}
        \end{array}\right.
\end{equation*}
where $P_n$ is the spectral projector of $H$ corresponding to the interval $[0,n]$ and ${\rm cl}(\mathcal{D}(H))$ is the closure of the domain of $H$. 
Note that $H$ does not need to be densely defined. If
$H$ is the Hamiltonian (energy observable) of a quantum system with underlying Hilbert space $\H$, then
$\Tr H\rho$ is the mean energy of a state $\rho$.

A positive semidefinite operator $H$ satisfies the \emph{Gibbs hypothesis} 
if
\begin{equation}\label{H-cond}
  \Tr\, e^{-\beta H}<+\infty\quad\textrm{for all}\;\,\beta>0.
\end{equation}
If this condition holds, then the von Neumann entropy is continuous on the set
\begin{equation*}
\C_{H,E}=\left\{\rho\in\S(\H)\,\vert\,\Tr H\rho\leq E\right\}
\end{equation*}
for any $E>0$. Its maximal value on this set is achieved by the \emph{Gibbs state}
\begin{equation}\label{Gibbs}
\gamma_H(E)\doteq e^{-\beta_H(E) H}/\Tr e^{-\beta_H(E) H},
\end{equation}
where the parameter $\beta_H(E)$ is determined by the equation $\Tr H e^{-\beta H}=E\Tr e^{-\beta H}$ \cite{W}.

We will use the function
\begin{equation}\label{F-def}
F_{H}(E)\doteq\sup_{\rho\in\C_{H,E}}S(\rho)=S(\gamma_H(E)).
\end{equation}
By Proposition 1 in \cite{EC}, the Gibbs condition (\ref{H-cond}) is equivalent to the asymptotic property
\begin{equation}\label{H-cond-a}
  F_{H}(E)=o\shs(E)\quad\textrm{as}\quad E\rightarrow+\infty.
\end{equation}

We will often assume that
\begin{equation}\label{star}
  E_0\doteq\inf\limits_{\|\varphi\|=1}\langle\varphi\vert H\vert\varphi\rangle=0.
\end{equation}

\section{Main results}\label{sec:Main}

Let $H$ be a positive operator on $\H$ satisfying conditions (\ref{H-cond}) and (\ref{star}) with the representation
\begin{equation*}
H=\sum_{i=0}^{+\infty} h_i|\tau_i\rangle\langle \tau_i|,\quad h_0=0,
\end{equation*}
and the domain $\mathcal{D}(H)=\{ \varphi\in\H_\mathcal{T}\,|\,\sum_{i=0}^{+\infty} h^2_i|\langle\tau_i|\varphi\rangle|^2<+\infty\}$, where
$\mathcal{T}=\left\{\tau_i\right\}_{i=0}^{+\infty}$ is the orthonormal system of eigenvectors of $H$
corresponding to the non-decreasing sequence $\left\{\smash{h_i}\right\}_{i=0}^{+\infty}$ of eigenvalues
tending to $+\infty$ and $\H_\mathcal{T}$ is the closure of the linear span of $\mathcal{T}$.

We will use the finite function
\begin{equation}\label{Z}
  Z_H(E) \doteq \Tr e^{-\beta_H(E)H}=\sum_{i=0}^{+\infty}e^{-\beta_H(E)h_i},
\end{equation}
where $\beta_H(E)$ is defined after (\ref{Gibbs}).  It is easy to see that
$Z_H(E)>Z_H(0)=\dim (\ker H)\geq 1$, and that $Z_H(E)$ is an increasing function on $\mathbb{R}_+$ (because $\beta_H(E)$ is a decreasing function).

Following \cite{BDJ}, we introduce the operator
\begin{equation*}
H_+ \doteq \sum_{i=0}^{+\infty} h_{i+1}|\tau_i\rangle\langle \tau_i|,
\end{equation*}
with  the domain $\mathcal{D}(H_+)=\{ \varphi\in\H_\mathcal{T}\,|\,\sum_{i=0}^{+\infty} h^2_{i+1}|\langle\tau_i|\varphi\rangle|^2<+\infty\}$, where
$\mathcal{T}=\left\{\tau_i\right\}_{i=0}^{+\infty}$  and $\H_\mathcal{T}$ is the linear span of $\mathcal{T}$.\smallskip

For each $E\geq h_1$, we write $F^+_{H}(E)$ for the function $F_{H_+}(E)$ defined in (\ref{F-def}) with $H=H_+$.
Since $H_+\geq H$, we have $F^+_{H}\leq F_{H}$. The precise relation between the functions $F^+_{H}$ and $F_{H}$ is established in the last claim of the following theorem.

\begin{theo}[{Semicontinuity bound for the von Neumann entropy of energy-constrained states}]\label{SCB} 
Let $H$ be a positive operator on $\H$  satisfying conditions (\ref{H-cond}) and (\ref{star}) and $E>0$.\smallskip

(A) If $\rho$ is a state in $\S(\H)$ such that $\Tr H\rho\leq E$ and $\eps\in(0,1]$ is arbitrary, then
\begin{equation}\label{SCB+}
\!   S(\rho)-S(\sigma)\leq\left\{\begin{array}{l}
       \!\varepsilon F^+_H(E/\varepsilon)+h(\varepsilon)\quad\;\, \textrm{if}\;\;  \varepsilon\in[0,a_H(E)]\\\\
        \!F_H(E)\qquad \qquad\qquad\textrm{if}\;\;  \varepsilon\in[a_H(E),1]
        \end{array}\right.,\quad   a_H(E)=1-1/Z_H(E),
        \end{equation}
for any state $\sigma$ in $\S(\H)$ such that $\,\frac{1}{2}\|\rho-\sigma\|_1\leq \varepsilon$, where the l.h.s. may be equal to $-\infty$. Here, $h(\varepsilon):=-\varepsilon \log(\varepsilon) - (1-\varepsilon)\log(1-\varepsilon)$ is the binary entropy, while $Z_H(E)$ is defined in (\ref{Z}).

\bigskip

(B) Semicontinuity bound (\ref{SCB+}) is optimal: for each $\varepsilon\in[0,1]$ there are states $\rho_\eps$ and $\sigma_\eps$ in $\S(\H)$
such that $\,\frac{1}{2}\|\rho_\eps-\sigma_\eps\|_1\leq\varepsilon$, $\Tr H\rho_\eps=E$ and $\Tr H\sigma_\eps=0$ for which the equality in (\ref{SCB+}) holds.

\bigskip

(C)
The function $F^+_{H}(E)$ in~\eqref{SCB+} satisfies the following identity:
\begin{equation}\label{F+def}
a_H(E)F^+_{H}(E/a_H(E))+h(a_H(E))=F_H(E).
\end{equation}
\end{theo}

\begin{rem}\label{main-r+}
The function $F_H^+(E)$ is only well defined for $E\geq h_1$. So, the term $\varepsilon F^+_H(E/\varepsilon)$  in (\ref{SCB+})
is not defined for $\varepsilon> E/h_1$. However, the right-hand side of (\ref{SCB+}) is always well defined (including the case $E/h_1<1$). This is because when $E<h_1,$ we have $\varepsilon\in[a(E),1]$ for any $\varepsilon> E/h_1$. This follows from the inequality
\begin{equation}\label{new-in}
a_H(E)\leq E/h_1 \text{ for }E <h_1.
\end{equation}
{The above condition can be verified directly for  a positive operator $H$ on $\H$  satisfying conditions (\ref{H-cond}) and (\ref{star}) with spectral gap $h_1>0$, which can be seen as follows:
\begin{equation}
\label{eq:bound}
    Z_H(E)=\Tr(e^{-\beta_H(E)H})=\Tr(\tfrac{H}{E} e^{-\beta_H(E)H})\ge\tfrac{h_1}{E} \sum_{i=1}^{\infty}e^{-\beta_H(E) h_i}=\frac{h_1(Z_H(E)-1)}{E},
\end{equation}
where the second equality follows from the definition of $\beta_H(E)$. Dividing the above inequality by $Z_H(E)$ and multiplying by $E/h_1$, shows that
\begin{equation*}
    a_H(E)= 1-\frac{1}{Z_H(E)}\le \frac{E}{h_1}.
\end{equation*}
}
\end{rem}

\noindent
Studying \eqref{eq:bound}, we see that the equality is obtained if and only if $H$ is a two-level Hamiltonian $H= h_0 \ket{0}\bra{0} + h_1 \sum_{i=1}^m \ket{i}\bra{i}$ with $h_0=0$ and $h_1>0$ and of arbitrarily but finite rank $m \ge 1$.

\begin{rem}\label{main-r}
Given that by monotonicity $F_H^+\leq F_H$ and that the operator $H$ satisfies condition (\ref{H-cond}), the equivalence between
 (\ref{H-cond}) and (\ref{H-cond-a}) implies that
\begin{equation}\label{H-cond-a+}
  F^+_{H}(E)=o\shs(E)\quad\textrm{as}\quad E\rightarrow+\infty.
\end{equation}
This property guarantees that the right-hand side of (\ref{SCB+}) tends to zero as $\eps\to0$.
\end{rem}

\noindent
Theorem \ref{SCB} provides the following continuity bound for the von Neumann entropy over the set of states with bounded energy in infinite-dimensional quantum systems. It improves on Theorem 6 of \cite{BDJ}, in which an analogous bound was obtained only for pairs of states that are sufficiently close to each other. This fully resolves the problem of finding an optimal continuity bound for the von Neumann entropy in infinite dimensions.

\begin{corr}[{Continuity bound for the von Neumann entropy of energy-constrained states}]\label{SCB-c}
Let $H$ be a positive operator on $\H$  satisfying conditions (\ref{H-cond}) and (\ref{star}). Let $\eps\in(0,1]$ be arbitrary. Then
\begin{equation}\label{SCB+c}
  \! |S(\rho)-S(\sigma)|\leq\left\{\begin{array}{l}
        \!\varepsilon F^+_H(E/\varepsilon)+h(\varepsilon)\quad\;\, \textrm{if}\;\;  \varepsilon\in[0,a_H(E)]\\\\
        \!F_H(E)\qquad \qquad\qquad\textrm{if}\;\;  \varepsilon\in[a_H(E),1]
        \end{array}\right.,\quad   a_H(E)=1-1/Z_H(E),
\end{equation}
for any states $\rho$ and $\sigma$ in $\S(\H)$ such that $\,\frac{1}{2}\|\rho-\sigma\|_1\leq \varepsilon$ and $\Tr H\rho,\Tr H\sigma\leq E$.

Moreover, the continuity bound (\ref{SCB+c}) is optimal: For each $\varepsilon\in[0,1]$, there are states $\rho_\eps$ and $\sigma_\eps$ in $\S(\H)$
such that $\,\frac{1}{2}\|\rho_\eps-\sigma_\eps\|_1=\varepsilon$, $\Tr H\rho_\eps=E$ and $\Tr H\sigma_\eps=0$ for which an equality in (\ref{SCB+c}) holds.
\end{corr}

\noindent
The right-hand side of (\ref{SCB+c}) tends to zero as $\eps\to0$ due to property (\ref{H-cond-a+}).\smallskip

\begin{example}\label{1-mode} Assume that $\H_1$ is a Hilbert space describing a one-mode quantum oscillator and  $H=N_1$, the number operator on $\H_1$, i.e.,
\begin{equation*}
N_1=\sum_{i=0}^{+\infty}i|\tau_i\rangle\langle \tau_i|,
\end{equation*}
where $\{{\ket{\tau_i}}\}_{i=0}^{+\infty}$ is the Fock basis in $\H_1$. It is easy to show that for any $E>0$,
$$
F_{N_1}(E)=g(E)\doteq(E+1)\log(E+1)-E\log E,
$$
and that the number operator $N_1$ satisfies conditions (\ref{H-cond}) and (\ref{star}).

Since $\beta_{N_1}(E)=\frac{d}{dE}F_{N_1}(E)=\log(1+1/E)$, we have
\begin{equation*}
Z_{N_1}(E)=\sum_{k=0}^{+\infty}e^{-\log(1+1/E)k}=1+E.
\end{equation*}
Hence $a_{N_1}(E)=E/(1+E)$.  It is not difficult to show that $F^+_{N_1}(E)=Eh(1/E)$ (this can be shown
directly or obtained as a solution of equation (\ref{F+def}) with $F_{N_1}(E)=g(E)$).

In this case, Theorem \ref{SCB} therefore implies the following semicontinuity bound:

\begin{corr}
If $\rho$ is a state in $\S(\H)$ such that $\Tr N_1\rho\leq E$ and $\eps\in(0,1]$ is arbitrary then
\begin{equation}\label{SCB+N}
   S(\rho)-S(\sigma)\leq\left\{\begin{array}{l}
       \!E h\!\left(\frac{\varepsilon}{E}\right)+h(\varepsilon)\;\;\;\, \textrm{if}\;\;  \varepsilon\in[0,\frac{E}{E+1}]\\\\
        \!g(E)\qquad \qquad\quad\,\textrm{if}\;\;  \varepsilon\in[\frac{E}{E+1},1]
        \end{array}\right.,
        \end{equation}
for any state $\sigma$ in $\S(\H)$ such that $\,\frac{1}{2}\|\rho-\sigma\|_1\leq \varepsilon$, where the l.h.s. of (\ref{SCB+N}) may be equal to $-\infty$.
\end{corr}
\smallskip

The semicontinuity bound (\ref{SCB+N}) is optimal for any $\eps\in[0,1]$. It agrees with the optimal continuity bound
 for the von Neumann entropy under the energy constraint induced by the operator $N_1$ presented in \cite[Theorem~5]{BDJ}.
\smallskip
\end{example}

\section{Some auxiliary results and a proof of Theorem~\ref{SCB}}\label{sec:proof}

\subsection{Preliminary results: Classical entropy bounds\label{sec:classRes}}

{In order to prove Theorem~\ref{SCB}, we begin by introducing two intermediate results: (i) a \emph{Fano-type} inequality for random variables with a {countably} infinite alphabet and an expectation value constraint (Theorem~\ref{theo:FanoG} below); (ii) a semicontinuity bound for entropies of random variables with a {countably} infinite alphabet and an expectation value constraint (Theorem~\ref{theo:classical} below). These results are extensions of Theorems~2 and~4 of~\cite{BDJ}, for the full parameter range of the error probability and the total variation distance, respectively.}


Let $(\Omega,\Sigma, \mathbb{P})$ be a probability space. Let $X,Y: \Omega\rightarrow   \mathbb N_0\coloneqq \mathbb N\cup\{0\}$ be discrete random variables with probabilities $p_X(i) \coloneqq \mathbb{P}(X=i)$ and $p_{XY}(i,j) \coloneqq \mathbb{P}(X=i,Y=j)$ for $i,j \in \mathbb N_0.$ 
\begin{defi}[Entropies: Discrete random variables]
\label{defi:disc_ent}
The Shannon entropy of $X$ is defined as
\begin{equation*}
    H(X) = -\sum_{i=0}^{\infty} p_X(i) \log(p_X(i)),
\end{equation*}
and the conditional entropy of $X$ given $Y$ is defined as
\begin{equation*}
    H(X\vert Y) = -\sum_{(i,j) \in \mathbb N_0^2} p_{XY}(i,j) \log(p_{XY}(i,j)/p_Y(i)).
\end{equation*}
\end{defi}

We consider a function $f : \mathbb N_0 \rightarrow \mathbb{R}_+$ with the following properties
\begin{equation} \label{eq:Boltzmann}
    \begin{aligned}
        (i) & \, f(0) = 0, \\
        (ii) & \, f \textrm{ is a non-decreasing function, and } \\
        (iii) & \, \textrm{for any } \lambda<0, \quad e^{-\lambda_0} \coloneqq \sum_{x \in \mathbb N_0} e^{\lambda f(x)} < \infty.
    \end{aligned}
\end{equation}

Consider the set of random variables with probability distributions $\{ p(x) \}_{x \in \mathbb N_0}$, subject to the constraint $\sum_{x \in \mathbb N_0} f(x) p(x) \le E$. It is known (see for instance~\cite[Section~3.4]{ETHln})
that the unique distribution achieving maximal Shannon entropy under this constraint is associated with the random variable $W_E$, which has the probability distribution $\{ w_{E}(x) \}_{x \in \mathbb N_0}$ defined by 
\begin{equation} \label{eq:defW1}
    w_E(x) \coloneqq e^{\lambda_0(E) + \lambda(E) f(x)}, \quad \forall x \in \mathbb N_0.
\end{equation}
{Here, $\lambda(E)$ (and hence also $\lambda_0(E)$) are determined by the conditions} 
\begin{equation*} 
    \sum_{x \in \mathbb N_0} w_{E}(x) = 1, \qquad \mathbb{E}[f(W_E)] \coloneqq \sum_{x \in \mathbb N_0} f(x) w_{E}(x) = E.
\end{equation*}
We refer to $W_E$ as the entropy-maximizing random variable for $\mathbb{E}[f(W_E)] = E$.

Similarly, define $\tilde{f} : \mathbb N_0 \rightarrow \mathbb{R}_+$ by
\begin{equation*} 
    \tilde{f}(x) = f(x+1).
\end{equation*}
Under this definition, $\tilde{f}$ also satisfies the properties $(ii)$ and $(iii)$ of~\eqref{eq:Boltzmann}, with $f$ replaced by $\tilde{f}$. This ensures the existence of the entropy-maximizing random variable $\widetilde{W}_E$ such that $\mathbb{E}[\tilde{f}(\widetilde{W}_E)] = E$.

\medskip

We then have the following extension of Theorem~2 of~\cite{BDJ}.

\begin{theo}[A Fano-type 
inequality for {random variables with a} countably infinite alphabet and a general constraint] \label{theo:FanoG}
Let $X$ and $Y$ be a pair of random variables taking values in $\mathbb N_0$, having a joint distribution $p_{XY}$, and satisfying the constraint $\mathbb{P}(X \ne Y)\leq\eps\in[0,1]$ and $\mathbb{E}\left[ f(X) \right] \le E$ for some $E>0$, where the function $f$ satisfies~\eqref{eq:Boltzmann}.
Then the following inequality holds:
\begin{equation} \label{F-theoG}
    H(X|Y) \leq
    \begin{cases}
        \eps H(\widetilde{W}_{E/\eps})+h(\eps) & \text{if} \quad \eps\in[0,a_f(E)], \\
        H( W_{E}) & \text{if} \quad \eps\in[a_f(E),1],
    \end{cases}
    \quad a_f(E):=1-e^{\lambda_0(E)}.
\end{equation}
\end{theo}

\begin{rem}\label{main-r+Class} 
The function $H(\widetilde{W}_{E})$ is only well defined for $E\geq f(1)$. Therefore, the term $\eps H(\widetilde{W}_{E/\eps})$ in~\eqref{F-theoG} 
is not defined for $\eps> E/f(1)$. However, the right-hand sides of~\eqref{F-theoG} 
are always well defined (including the case $E/f(1)<1$). This is because when $E<f(1),$ we have $\eps \in [a_f(E),1]$ for any $\eps > E/f(1)$. This follows from the inequality
\begin{equation}\label{new-inclass}
a_f(E)\leq E/f(1) \quad\text{for}\quad E < f(1),
\end{equation}
which is a simple adaptation of~\eqref{new-in} in Remark~\ref{main-r+} to the classical case. The above argument also makes the right-hand side of the inequality~\eqref{SCB++c} below well defined, even though there too
the term $\eps H(\widetilde{W}_{E/\eps})$ 
is not defined for $\eps> E/f(1)$. \end{rem}

\begin{rem}\label{rem:tightClass-1}
The Fano-type inequality~\eqref{F-theoG} is optimal for $\eps\in[0,1]$:
for each $\eps\in[0,1]$, there are random variables $X_\eps$ and $Y_\eps$ such that $\,\mathbb{P}(X_\eps \ne Y_\eps)\leq \eps$ and $\,\mathbb E(f(X_\eps))=E$ for which equality in~\eqref{F-theoG} holds.
\end{rem}

\begin{proof}[Proof of Remark~\ref{rem:tightClass-1}]
Assume first that $\eps\in[0,a_f(E)]$.
From inequality~\eqref{new-inclass}, this assumption  implies $\eps\in(0, \min\{1, E/f(1)\}]$. Since $E/\varepsilon\geq f(1)$, the random variable $\widetilde{W}_{E/\eps}$ is well defined.
Consider the random variables $(X_\eps,Y_\eps)$ characterized by the joint probability distribution $p_{X_\eps Y_\eps}$ which is defined as follows:
\begin{equation} \label{eq:probaTightb}
    p_{X_\eps Y_\eps}(n,m) \coloneqq \begin{cases}
    1 - \varepsilonilon & \quad \mathrm{if} \, n = 0, m = 0, \\
    \varepsilonilon \, \tilde{w}_{E/\eps}(n-1) & \quad \forall \, n \in \mathbb N, \, \mathrm{for} \, m = 0, \\
    0, & \quad \mathrm{else},
    \end{cases}
\end{equation}
where $\{\tilde{w}_{E/\eps}(n)\}_{n \in \mathbb N_0}$ is the probability distribution of the entropy-maximizing random variable $\widetilde{W}_{E/\eps}$ satisfying $\mathbb{E}[\tilde{f}(\widetilde{W}_{E/\eps})] = E/\eps$.
First note that
\begin{equation}\label{eq:probaTightb2}
        \mathbb{E}(f(X_\eps))
        = \sum_{n \in \mathbb N_0} f(n) p_{X_\eps Y_\eps}(n,0)
        = \varepsilon \sum_{n \in \mathbb N_0} f(n+1) \tilde{w}_{E/\eps}(n)  = \varepsilon \mathbb{E}[\tilde{f}(\widetilde{W}_{E/\eps})]  = E.
\end{equation}
Secondly, note that
\begin{equation*}
    \mathbb{P}(X_\eps \ne Y_\eps)  = 1 - \mathbb{P}(X_\eps = Y_\eps)  = 1 - \sum_{n \in \mathbb N_0} p_{X^*Y^*}(n,n)  = \varepsilon.
\end{equation*}
Finally, since $H(Y^*) =0$,
\begin{equation}\label{eq:probaTightb3}
    H(X_\eps|Y_\eps) = H(X_\eps Y_\eps)
    = h(\varepsilon) + \varepsilon H(\widetilde{W}_{E/\eps}).
\end{equation}

If $\eps\in(a_f(E),1]$ then take the random variables $X_\eps$ and $Y_\eps$ characterized  by the joint probability distribution   $p_{X_{\eps_*}Y_{\eps_*}}$  defined in (\ref{eq:probaTightb})  with $\eps_*=a_f(E)$. Then $\mathbb{E}(X_\eps) = E$,  $\mathbb{P}(X_\eps \ne Y_\eps)=\eps_*\leq \eps$  and $H(X_\eps|Y_\eps) = H(W_{E})$.  The last equality is a corollary of the identity 
\begin{equation}\label{last-f}
a_f(E)H(\widetilde{W}_{E/a_f(E)})+h(a_f(E))=H(W_{E}),
\end{equation}
which follows from the proof of Theorem~\ref{theo:FanoG} (
with $a_*=a_f(E)$) presented in Section \ref{sec:proofC1}.
\end{proof}

\medskip

We also have the following extension of Theorem~4 of~\cite{BDJ}.

\begin{theo}[Constrained semicontinuity bound for the Shannon entropy of random variables with countably infinite alphabets] \label{theo:classical}
Let $X$ and $Y$ be a pair of random variables taking values in $\mathbb N_0$ such that $\Vert X-Y\Vert_{\operatorname{TV}}\leq\eps$ and $\mathbb E(f(X)) \le E$  for some $E>0$, where the function $f$ satisfies~\eqref{eq:Boltzmann}, then 

\begin{equation} \label{SCB++c}
    H(X)-H(Y) \leq
    \begin{cases}
        \eps H(\widetilde{W}_{E/\eps})+h(\eps) & \text{if} \quad \eps\in[0,a_f(E)], \\
        H( W_{E}) & \text{if} \quad \eps\in[a_f(E),1],
    \end{cases}
    \quad a_f(E):=1-e^{\lambda_0(E)}.
\end{equation}
\end{theo}


\begin{rem}\label{rem:tightClass-2}
The semicontinuity bound~\eqref{SCB++c} is optimal for $\eps\in[0,1]$:
For each $\eps\in[0,1]$, there are random variables $X_\eps$ and $Y_\eps$ such that $\Vert X_\eps-Y_\eps\Vert_{\operatorname{TV}} \leq\eps$, $\mathbb E(f(X_\eps))=E$ for which equality in~\eqref{SCB++c} holds.
\end{rem}

\begin{proof}[Proof of Remark~\ref{rem:tightClass-2}]
Assume that $\eps\in[0,a_f(E)]$. Consider the random variables $X_\eps$ and $Y_\eps$ characterized by the probability distributions $p_{X_\eps}$ which is defined as follows:
\begin{equation*} 
    p_{X_\eps}(n) \coloneqq \begin{cases}
    1 - \eps & \quad \mathrm{if} \, n = 0, \\
    \eps \, \tilde{w}_{E/\eps}(n-1) & \quad \mathrm{else},
    \end{cases}
\end{equation*}
and $p_{Y_\eps}$ which is defined as follows:
\begin{equation}\label{Y-d}
    p_{Y_\eps}(m) \coloneqq \begin{cases}
    1 & \quad \mathrm{if} \, m = 0, \\
    0 & \quad \mathrm{else}.
    \end{cases}
\end{equation}
Note that $p_{X_\eps}$ and $p_{Y\eps}$ simply correspond to the two marginals of $p_{X_\eps Y_\eps}$ defined in~\eqref{eq:probaTightb}. From this and from \eqref{eq:probaTightb2}, we know that $\mathbb{E}(X_\eps) = E$. From \eqref{eq:probaTightb3}, we have $H(X_\eps) = h(\varepsilon) + \varepsilon H(\widetilde{W}_{E/\eps})$. Obviously, we also have  $H(Y_\eps) = 0$. Finally, it is easy to see that $\Vert X_\eps-Y_\eps\Vert_{\operatorname{TV}} = \eps$.

If $\eps\in(a_f(E),1]$ then take the random variables $X_\eps$ and $Y_\eps$ characterized  by the probability distributions $w_{E}$ and  $p_{Y_{\eps_*}
}$  defined, respectively, in  (\ref{eq:defW1}) and (\ref{Y-d}). Then $\mathbb{E}(X_\eps) = E$,  $H(X_\eps) = H(W_{E})$ and $H(Y_\eps) = 0$. It is easy to see that $\Vert X_\eps-Y_\eps\Vert_{\operatorname{TV}} = a_f(E)\leq \eps$.
\end{proof}

 A continuity bound for the Shannon  entropy 
of random variables with countably infinite alphabets follows immediately as a corollary of Theorem~\ref{theo:classical}:

\begin{corr}[Continuity bound for the Shannon  entropy 
of random variables with countably infinite alphabets] \label{Sh-CB}
Let $X$ and $Y$ be a pair of random variables taking values in $\mathbb N_0$ such that $\Vert X-Y\Vert_{\operatorname{TV}}\leq\eps$ and $\mathbb E(f(X)),\mathbb E(f(Y)) \le E$  for some $E>0$, where the function $f$ satisfies~\eqref{eq:Boltzmann}, then 

\begin{equation} \label{Sh-CB+}
    |H(X)-H(Y)| \leq
    \begin{cases}
        \eps H(\widetilde{W}_{E/\eps})+h(\eps) & \text{if} \quad \eps\in[0,a_f(E)], \\
        H( W_{E}) & \text{if} \quad \eps\in[a_f(E),1],
    \end{cases}
    \quad a_f(E):=1-e^{\lambda_0(E)}.
\end{equation}
\end{corr}

The  example 
considered in the proof of Remark \ref{rem:tightClass-2} also allows us to prove the following: 

\begin{rem}\label{rem:tightClass-3}
The continuity bound~\eqref{Sh-CB+} is optimal for $\eps\in[0,1]$:
for each $\eps\in[0,1]$, there are random variables $X_\eps$ and $Y_\eps$ such that $\Vert X_\eps-Y_\eps\Vert_{\operatorname{TV}} \leq\eps$, $\mathbb E(f(X_\eps))=E$, $\mathbb E(f(Y_\eps))=0$ for which equality in~\eqref{Sh-CB+} holds.
\end{rem}

\subsection{Proof of Theorem~\ref{theo:FanoG}\label{sec:proofC1}}

The first part of the proof is exactly the same as that of Theorem~2 of~\cite{BDJ}. However, since the proof of the latter was only explicitly written for the particular case where $f$ is the identity function, we reproduce its generalization here for completeness.

Define the set \( Z \coloneqq \{ m \in \mathbb{N}_0 : p_{XY}(m,m) \geq p_{XY}(0,m) \} \) and a new random variable \( X' \) such that $X'Y$ has joint probability distribution \( p_{X'Y} \) specified as follows. For each \( m \in \mathbb{N}_0 \),
\begin{equation} \label{eq:Xp1}
    p_{X'Y}(0,m) \coloneqq \begin{cases}
    p_{XY}(m,m), & \text{if } m \in Z, \\
    p_{XY}(0,m), & \text{otherwise},
    \end{cases}
\end{equation}
and for \( m \in \mathbb{N} \),
\begin{equation} \label{eq:Xp2}
    p_{X'Y}(m,m) \coloneqq \begin{cases}
    p_{XY}(0,m), & \text{if } m \in Z, \\
    p_{XY}(m,m), & \text{otherwise}.
    \end{cases}
\end{equation}
For all other cases, set \( p_{X'Y}(n,m) \coloneqq p_{XY}(n,m) \).

We begin by noting that \( \mathbb{E}(f(X')) \leq \mathbb{E}(f(X)) \leq E \), as follows, where we use the fact that $f(0)=0$: 
\begin{equation} \label{upxpb}
    \begin{aligned}
        \mathbb{E}(f(X')) &= \sum_{n \in \mathbb{N}_0} f(n) \sum_{m \in \mathbb{N}_0} p_{X'Y}(n,m) \\
        &= \sum_{n \in Z} f(n) \, p_{X'Y}(n,n) + \sum_{n \in \mathbb{N}_0 \setminus Z} f(n) \, p_{X'Y}(n,n) + \sum_{n \in \mathbb{N}_0} f(n) \sum_{\substack{m \in \mathbb{N}_0 \\ m \ne n}} p_{X'Y}(n,m) \\
        &= \sum_{n \in Z} f(n) \, p_{X'Y}(n,n) + \sum_{n \in \mathbb{N} \setminus Z} f(n) \, p_{X'Y}(n,n)  + \sum_{n \in \mathbb{N}} f(n) \sum_{\substack{m \in \mathbb{N}_0 \\ m \ne n}} p_{XY}(n,m) \\
        &= \sum_{n \in Z} f(n) \, p_{XY}(0,n) + \sum_{n \in \mathbb{N}_0 \setminus Z} f(n) \, p_{XY}(n,n) + \sum_{n \in \mathbb{N}} f(n) \sum_{\substack{m \in \mathbb{N}_0 \\ m \ne n}} p_{XY}(n,m) \\
        &\leq \sum_{n \in Z} f(n) \, p_{XY}(n,n) + \sum_{n \in \mathbb{N}_0 \setminus Z} f(n) \, p_{XY}(n,n) + \sum_{n \in \mathbb{N}_0} f(n) \sum_{\substack{m \in \mathbb{N}_0 \\ m \ne n}} p_{XY}(n,m) \\
        &= \mathbb{E}(f(X)) \leq E,
    \end{aligned}
\end{equation}
where the first inequality arises from the definition of \( Z \) and the second follows from the constraint \( \mathbb{E}(f(X)) \leq E \).

Next, observe that \( H(X'|Y) = H(X|Y) \) since \( H(X'Y) = H(XY) \), which is straightforward to verify. 
Given that conditioning reduces entropy, we have
\begin{equation} \label{eqfb}
     H(X|Y) = H(X'|Y) \leq H(X').
\end{equation}
Thus, it remains to find an upper bound on \( H(X') \).

Let \( \varepsilon' \coloneqq 1 - p_{X'}(0) \), and note that
\begin{equation*}
    \begin{aligned}
        \varepsilon' &= 1 - \sum_{m \in \mathbb{N}_0} p_{X'Y}(0,m) \\
        &= 1 - \sum_{m \in Z} p_{X'Y}(0,m) - \sum_{m \in \mathbb{N}_0 \setminus Z} p_{X'Y}(0,m) \\
        &= 1 - \sum_{m \in Z} p_{XY}(m,m) - \sum_{m \in \mathbb{N}_0 \setminus Z} p_{XY}(0,m).
    \end{aligned}
\end{equation*}
Since \( p_{XY}(m,m) \geq p_{XY}(0,m) \) if and only if \( m \in Z \), it follows that \( p_{XY}(m,m) < p_{XY}(0,m) \) for all \( m \in \mathbb{N}_0 \setminus Z \). Thus,
\begin{equation} \label{eq:epsp}
    \begin{aligned}
        \varepsilon' &\leq 1 - \sum_{m \in Z} p_{XY}(m,m) - \sum_{m \in \mathbb{N}_0 \setminus Z} p_{XY}(m,m) \\
        &= 1 - \sum_{m \in \mathbb{N}_0} p_{XY}(m,m) = \mathbb{P}(X \ne Y) = \varepsilon.
    \end{aligned}
\end{equation}
Hence, we find that
\[
\sum_{n=1}^{\infty} p_{X'}(n) = \varepsilon' \leq \varepsilon.
\]
Thus,
\begin{equation*}
    \begin{aligned}
        H(X') &= -(1 - \varepsilon') \log(1 - \varepsilon') - \sum_{n=1}^{\infty} p_{X'}(n) \log p_{X'}(n) \\
	&= h(\varepsilon') + \varepsilon' \log \varepsilon' -  \sum_{n=1}^{\infty} p_{X'}(n) \log p_{X'}(n) \\
	&= h(\varepsilon') - \sum_{n=1}^{\infty} p_{X'}(n) \log \frac{p_{X'}(n)}{\varepsilon'} \\
	&= h(\varepsilon') - \varepsilon' \sum_{n=0}^{\infty} r(n) \log r(n) = h(\varepsilon') + \varepsilon' H(R),
    \end{aligned}
\end{equation*}
where \( R \) is a random variable on \( \mathbb{N}_0 \) with distribution
\[
	 r(n):=\mathbb{P}(R = n) \coloneqq \frac{p_{X'}(n+1)}{\varepsilon'}, \quad \forall n \in \mathbb{N}_0.
\]
To bound \( H(R) \), we estimate \( \mathbb{E}(\tilde{f}(R)) \) using \eqref{upxpb}:
\begin{align*}
  \mathbb{E}(\tilde{f}(R)) &= \frac{1}{\varepsilon'} \sum_{n = 0}^{\infty} \tilde{f}(n) p_{X'}(n+1) = \frac{1}{\varepsilon'} \sum_{n = 0}^{\infty} {f}(n) p_{X'}(n)= \frac{\mathbb{E}({f}(X'))}{\varepsilon'}  \le \frac{E}{\varepsilon'}.
\end{align*}
This shows that 
\[ H(X') \le h(\varepsilon') + \varepsilon' H(\widetilde{W}_{E/\varepsilon'})\]
Since $\varepsilon' \le \varepsilon$ and $\varepsilon' \le E/f(1)$, by  
\[ E \ge \mathbb E(f(X')) = \sum_{n \in \mathbb N} f(n) p_{X'}(n) \ge f(1) \sum_{n \in \mathbb N}  p_{X'}(n) = f(1) \varepsilon', \]
we have 
\[ H(X') \le \max_{\delta \in [0,c(\eps)]} ( \delta H(\widetilde{W}_{E/\delta}) + h(\delta)), \quad \text{with} \quad c(\eps) \coloneqq \min \{ \eps,E/f(1) \}.\]
By combining this estimate with~\eqref{eqfb}, we thus have that
\begin{equation*} 
    H(X \vert Y) \le \max_{\delta \in [0,c(\eps)]} ( \delta H(\widetilde{W}_{E/\delta}) + h(\delta)).
\end{equation*}

Defining $F_f(x):=  H(\widetilde{W}_{x})$   for $x\in(f(1),+\infty)$ and
$G_E(\delta) \coloneqq \delta F_f(E/\delta)+h(\delta)$
for {$\delta \in (0,\min\{1, E/f(1)\}),$} then 
\[G_E'(\delta) = F_f(E/\delta)-\frac{E F_f'(E/\delta)}{\delta} + \log\frac{1-\delta}{\delta},\qquad G_E''(\delta) = \frac{E^2 F_f''(E/\delta)}{\delta^3} - \frac{1}{\delta-\delta^2}.\]
As in the quantum case (using~\cite[Proposition~1]{EC}), we have $F_f'' \le 0$  on $(f(1),+\infty)$ and  $\lim_{x\to f(1)^+}F_f'(x)=+\infty$. Thus, we find that $G_E$ is concave on $(0,\min\{1, E/f(1)\}]$ and $G_E'(\delta)<0$ for all $\delta$ sufficiently close to $\min\{1, E/f(1)\}$. This implies the existence 
of a number $a_*$  in $(0,\min\{1, E/f(1)\})$  for which $G_E$ achieves its maximum value. It follows that
\begin{equation*}
        g_E(\varepsilon) \coloneqq \max_{\delta\in[0,c(\varepsilon)]}G_E(\delta)
    = \begin{cases}
        G_E(\varepsilon) & \text{if} \quad \eps \in [0,a_*] \\
        G_E(a_{*}) & \text{if} \quad \eps \in [a_*,1].
    \end{cases}
\end{equation*}

Next, we analyze the function
\begin{equation*}
    \Delta_E(\eps) \coloneqq \sup \Big\{H(X|Y)\,| X,Y: \Omega\rightarrow \mathbb N_0, \mathbb E(f(X)) \le E, P(X \neq Y) \leq\eps\Big\},
\end{equation*}
for $\eps \in [0,1]$ and observe that $\Delta_E(\eps)$ is a non-decreasing function on $[0,1]$.
It is clear that $\Delta_E(\varepsilon)\le \max_{\substack{X:\Omega \to \mathbb N_0; \\ \mathbb E(f(X))\le E}} H(X) = H(W_{E}).$ 
In addition, let $X = W_E$ and $Y=0$, then the event $X \neq Y$ has probability $a_f(E):=1-e^{\lambda_0(E)}.$ Thus, for $\varepsilon \in [a_f(E),1]$ we have equality $\Delta_E(\eps) =H(W_{E}).$

Furthermore, $\Delta_E(\eps) < H(W_E)$ when $\eps < a_f(E)$. To show this, for any random variables $X,Y: \Omega\rightarrow \mathbb N_0$, consider the random variable $X': \Omega\rightarrow \mathbb N_0$ with distribution $p_{X'}$ being the marginal of the joint distribution defined in~\eqref{eq:Xp1} and~\eqref{eq:Xp2}, with $\mathbb{E}(f(X')) \leq E$ (see~\eqref{upxpb}). Moreover, consider the trivial random variable $Y': \Omega\rightarrow \mathbb N_0$ with distribution $p_{Y'}$ defined as $p_{Y'}(y) = 1$ if $y=0$ and $p_{Y'}(y) = 0$ else. We have $P(X' \neq Y') = 1 - p_{X'}(0) = \eps' \leq \eps$ (see~\eqref{eq:epsp}). From~\eqref{eqfb}, we also have $H(X|Y) \leq H(X') = H(X'|Y')$. Taking all these considerations into account, we can write
\begin{equation}\label{D-sup}
    \Delta_E(\eps) = \sup \Big\{H(X)\,| X: \Omega\rightarrow \mathbb N_0, \mathbb E(f(X)) \le E, p_X(0) \geq 1-\eps \Big\}.
\end{equation}
Note than the supremum in (\ref{D-sup}) is attained for any $\eps\in[0,1]$ because of the continuity of the Shannon entropy on  the set of probability distributions  over which the supremum is taken (see~\cite[Section~3.4]{ETHln}) and due to the compactness (w.r.t. the $\ell_1$ norm) of this set (by the classical analogue of the Lemma in \cite{H-c-w-c}). 

Now, when $\eps < a_f(E)$, the supremum in $\Delta_E(\eps)$, as rewritten in the above equation, is taken over all variables $X$ satisfying $\mathbb E(f(X)) \leq E$ and $p_{X}(0) > 1-a_f(E) = e^{\lambda_0(E)}$. The supremum
\begin{equation*}
    \sup \Big\{H(X)\,| X: \Omega\rightarrow \mathbb N_0, \mathbb E(f(X)) \le E \Big\}
\end{equation*}
(equal to $H(W_E)$) is uniquely reached for $X_*=W_E$, which satisfies $p_{W_E}(0) = e^{\lambda_0(E)}$ (see~\cite[Section~3.4]{ETHln}). Since the random variable that reaches the supremum is unique, we necessarily have that  $\Delta_E(\eps) < H(W_E)$.

By our general bound, we have $\Delta_E(\varepsilon) \le g_E(\varepsilon)$ for all $\varepsilon \in [0,1]$ and $\Delta(\varepsilon)=g_E(\varepsilon)$ for all $\varepsilon \in [0,a_*]$ by the same example as in Remark~\ref{rem:tightClass-1}. If $a_f(E)< a_*$, then for $\varepsilon \in [a_f(E),a_*]$ we have $\Delta(\eps) =H(W_{E}) =g_E(\varepsilon)$ which is a contradiction, as $g_E(\varepsilon)$ is not constant.  If $a_*<a_f(E)$, then 
the properties of the function $\Delta_E$ derived above imply $g_E(a_*)=\Delta_E(a_*)<\Delta_E(a_f(E))=H(W_{E})$, which is a contradiction, as $\Delta_E(\eps)\leq g_E(\eps)\leq g_E(a_*)$ for all $\varepsilon\in[0,1]$, by definition of $a_*$. Thus, we conclude that $a_*=a_f(E)$.

\subsection{Proof of Theorem~\ref{theo:classical}\label{sec:proofC2}}

Following the approach in the proof of Theorem~3 from~\cite{Sason2013}, let \((\hat{X}, \hat{Y})\) denote a maximal coupling of \((X, Y)\).
Define the function $\kappa_E(\eps)$ to be the right-hand side of~\eqref{F-theoG}.
Noting that \(H({X}) = H({\hat{X}})\) and \(H({Y}) = H({\hat{Y}})\), we obtain
\begin{equation*}
    \begin{aligned}
         H({X}) - H({Y})  & =  H({\hat{X}}) - H({\hat{Y}})  \\
	& =  H({\hat{X}}\vert{\hat{Y}}) - H({\hat{Y}}\vert{\hat{X}})  \\
	& \leq  H({\hat{X}}\vert{\hat{Y}}) \\
	& \stackrel{(1)}{\leq}  \kappa_E(\eps') \\
	& \stackrel{(2)}{=} \kappa_E(\eps)
    \end{aligned}
\end{equation*}
where \(\varepsilon' \coloneqq \mathbb{P}(\hat{X} \ne \hat{Y})\). Here, \((1)\) follows from Theorem~\ref{theo:FanoG}, while \((2)\) uses the fact that for a maximal coupling \((\hat{X}, \hat{Y})\) of \((X, Y)\),
\begin{equation*}
    \mathbb{P}(\hat{X} \neq \hat{Y}) = \Vert X - Y \Vert_{\operatorname{TV}}.
\end{equation*}
For random variables with finite state spaces, a proof of the above identity appears in \cite{Sason2013}. The extension to infinite-state alphabets follows in a similar way.

\subsection{Proof of Theorem~\ref{SCB}\label{sec:proofsubMain}}
\smallskip

\noindent
A) 
Let the spectral decompositions of $\rho$ and $\sigma$ be given by
\begin{equation*}
    \rho = \sum_{n=0}^{+\infty} r(n) \proj{\phi_n}, \qquad \sigma = \sum_{n=0}^{+\infty} s(n) \proj{\psi_n}.
\end{equation*}
Consider the passive states
\begin{equation*}
    \rho^{\downarrow} = \sum_{n=0}^{+\infty} r^{\downarrow}(n) \proj{n}, \qquad \sigma^{\downarrow} = \sum_{n=0}^{+\infty} s^{\downarrow}(n) \proj{n},
\end{equation*}
where the states $\ket{n}$ for $n = 0, 1, \cdots$ are eigenstates of the Hamiltonian $\ha$. Here, $\{r^{\downarrow}(n)\}_{n \in \mathbb N_0}$ represents the distribution containing the non-zero elements of $\{r(n)\}_{n \in \mathbb N_0}$ arranged in non-increasing order, that is, $r^{\downarrow}(n) \geq r^{\downarrow}(n+1)$ for all $n \in \mathbb N_0$, and similarly for $\{s^{\downarrow}(n)\}_{n \in \mathbb N_0}$.
We obviously have that $S(\rho^{\downarrow}) = S(\rho)$ and $S(\sigma^{\downarrow}) = S(\sigma)$, so that
\begin{equation*}
	S(\rho) - S(\sigma) = S(\rho^{\downarrow}) - S(\sigma^{\downarrow}).
\end{equation*}
Furthermore, from the Courant-Fischer theorem~\cite[Proposition II-3]{BDJ}, we have $\tr(\ha \rho^{\downarrow}) \leq \tr(\ha \rho) \leq E$, $\tr(\ha \sigma^{\downarrow}) \leq \tr(\ha \sigma) \leq E$, while Mirsky’s inequality (\cite{Mirsky}, see also \cite[(1.22)]{Mirsky-rr} for a version in infinite dimensions) implies that
\begin{equation*}
    \eps' \coloneqq \frac{1}{2} ||\rho^{\downarrow}-\sigma^{\downarrow}||_1 \leq \frac{1}{2} ||\rho-\sigma||_1 \leq \varepsilon.
\end{equation*}
Let $X$ and $Y$ denote random variables on $\mathbb N_0$, with probability distributions $r \equiv \{r^{\downarrow}(n)\}_{n \in \mathbb N_0}$ and $s \equiv \{s^{\downarrow}(n)\}_{n \in \mathbb N_0}$, respectively. Then ${\mathbb{E}}(f(X)) = \Tr(\ha \rho^{\downarrow}) \leq E$, where we have defined the function $f : \mathbb N_0 \rightarrow \mathbb{R}_+$ as $f(x) = h_x$ for all $x \in N_0$. Note that since the Hamiltonian $H$ satisfies the Gibbs hypothesis~\eqref{H-cond}, the function $f$ satisfies~$(iii)$ in~\eqref{eq:Boltzmann}. Moreover, $f$ also satisfies~$(i)$ and~$(ii)$ in~\eqref{eq:Boltzmann} since $H$ is a grounded Hamiltonian.
Furthermore, $H(X) = S(\rho^{\downarrow})$, $H(Y) = S(\sigma^{\downarrow})$ and $\Vert X-Y\Vert_{\operatorname{TV}} = \eps' \leq \eps$.
So, by applying  Theorem~\ref{theo:classical} we obtain
\begin{equation*}
    S(\rho) - S(\sigma)\leq H(X) - H(Y) \leq \kappa_E(\eps')\leq \kappa_E(\eps),
\end{equation*}
where  $\kappa_E(\eps)$ to be the right-hand side of~(\ref{SCB+}). The last
inequality is due to the monotonicity of $\kappa_E(\eps)$ as a function of $\varepsilon$.

B) The optimality of the semicontinuity bound ~(\ref{SCB+}) can be shown by using Remark \ref{rem:tightClass-2}. Indeed,  
it suffices to take the states $\rho_{\eps}$ and $\sigma_{\eps}$
to be diagonal in the basis of eigenvectors of the operator $H$
whose spectra (arranged in non-increasing order) coincide
with the probability distributions of the random variables $X_\eps$ and $Y_\eps$ from Remark \ref{rem:tightClass-2}.

C) 
The identity (\ref{F+def}) is a quantum version of the identity (\ref{last-f}). This completes the proof of Theorem~\ref{SCB}.

\medskip

Our second main result, namely, a continuity bound
for the von Neumann entropy of energy-constrained systems, stated in Corollary~\ref{SCB-c}, follows immediately from Theorem~\ref{SCB-c}.

\section*{Acknowledgments}

MGJ acknowledges support from the Fonds de la Recherche Scientifique – FNRS (Belgium).

\bibliographystyle{IEEEtran}
\bibliography{BiblioInfo}

\end{document}